\documentclass[12pt]{iopart}

\usepackage{graphicx}
\usepackage{amssymb}

\usepackage{color}
\definecolor{mygreen}{rgb}{0,0.5,0}
\definecolor{myblue}{rgb}{0,0,0.75}
\definecolor{mymagenta}{cmyk}{0,1,0,0.12}

\usepackage{umoline}

\begin{document}

\title{Frequency measurement of the $^{1}\mathrm{S}_{0}, F=5/2\leftrightarrow\,^{3}\mathrm{P}_{1}, F=7/2$ transition of $^{27}$Al$^{+}$ via quantum logic spectroscopy with $^{40}$Ca$^{+}$}

\author{M.~Guggemos$^{1,2}$\ddag, M.~Guevara-Bertsch$^{1,2}$\footnote{These authors contributed equally to this work.}, D.~Heinrich$^{1,2}$, O.~A.~Herrera-Sancho$^{1,3,4,5}$, Y.~Colombe$^{2}$, R.~Blatt$^{1,2}$, C.~F.~Roos$^{1,2}$}

\address{$^1$ Institut f\"ur Quantenoptik und Quanteninformation, \"Osterreichische Akademie der Wissenschaften, Technikerstr. 21a, 6020 Innsbruck, Austria}
\address{$^2$ Institut f\"ur Experimentalphysik, Universit\"at Innsbruck, Technikerstr. 25, 6020 Innsbruck, Austria}
\address{$^3$ Escuela de F\'isica, Facultad de Ciencias, Universidad de Costa Rica, 2060 San Pedro, San Jos\'e, Costa Rica.}
\address{$^4$ Centro de Investigaci\'on en Ciencia e Ingenier\'ia de Materiales, Universidad de Costa Rica, 2060 San Pedro, San Jos\'e, Costa Rica}
\address{$^5$ Centro de Investigaci\'on en Ciencias At\'omicas Nucleares y Moleculares, Universidad de Costa Rica, 2060 San Jos\'e, Costa Rica}
\ead{christian.roos@uibk.ac.at}
\vspace{10pt}

\begin{abstract}
We perform quantum logic spectroscopy with a $^{27}$Al$^{+}$/$^{40}$Ca$^{+}$ mixed ion crystal in a linear Paul trap for a measurement of the $(3s^{2})\,^{1}\mathrm{S}_{0} \leftrightarrow \, (3s3p)\,^{3}\mathrm{P}_{1}, F=7/2$ intercombination transition in $^{27}$Al$^{+}$. Towards this end, Ramsey spectroscopy is used for probing the transition in $^{27}$Al$^{+}$ and the $(4s^{2})\,\mathrm{S}_{1/2} \leftrightarrow \, (4s3d)\,\mathrm{D}_{5/2}$ clock transition in $^{40}$Ca$^{+}$ in interleaved measurements. By using the precisely measured frequency  of the clock transition in $^{40}$Ca$^{+}$  as a frequency reference, we determine the frequency of the intercombination line to be $\nu_{^{1}\mathrm{S}_{0} \leftrightarrow \,^{3}\mathrm{P}_{1},F=7/2}$=1122\,842\,857\,334\,736(93)\,Hz and the Land\'e g-factor of the excited state to be $g_{^{3}\mathrm{P}_{1}, F=7/2}$=0.428132(2). 

\end{abstract}

%

\section{Introduction}

Narrow atomic transitions in the optical frequency domain can serve as extremely accurate frequency references with the potential to determine the atomic transition frequency with an uncertainty as low as one part in $10^{18}$ \cite{Dehmelt:1982}. Over the last decade, optical atomic clocks \cite{Ludlow:2015} probing narrow transitions in neutral atoms or trapped ions have been developed that are rapidly approaching this ambitious goal. Through the interrogation of a large number of atoms, frequency standards based on neutral atoms held in optical lattices, achieve excellent short-term instabilities \cite{Nemitz:2016,Schioppo:2017}. Frequency standards interrogating only a single ion suffer from larger short-term instabilities but have the advantage to facilitate the control over systematic frequency shifts. Both, single-ion and lattice clocks, have demonstrated a control of systematic frequency uncertainties at the level of $10^{-18}$ \cite{Oelker:2019A, Brewer:2019A,Huntemann:2016,Chen:2017,hankin2019systematic}. Thus, they surpass the performance of Cs clocks as testified by frequency ratio measurements between different clocks with an uncertainty of about $5\cdot 10^{-17}$ \cite{Nemitz:2016,Rosenband:2008}. 

Single-ion frequency standards based on Hg$^+$, Ca$^+$, Sr$^+$, or Yb$^+$ have demonstrated systematic uncertainties below $10^{-16}$ \cite{Huntemann:2016,Oskay:2006,Barwood:2014,Huang:2016}. These ions species are attractive since relativistic frequency shifts can be made small by laser-cooling the ions to sub-millikelvin temperatures and their clock transition can be measured using the electron shelving technique, which enables the detection of an ion's quantum state with an efficiency close to 100\%. They require, however, a careful evaluation of systematic frequency shifts such as electric quadrupole, ac-Stark or black-body radiation-induced shifts.

\begin{figure}[tb]
\begin{center}
\includegraphics[width=1.0\textwidth]{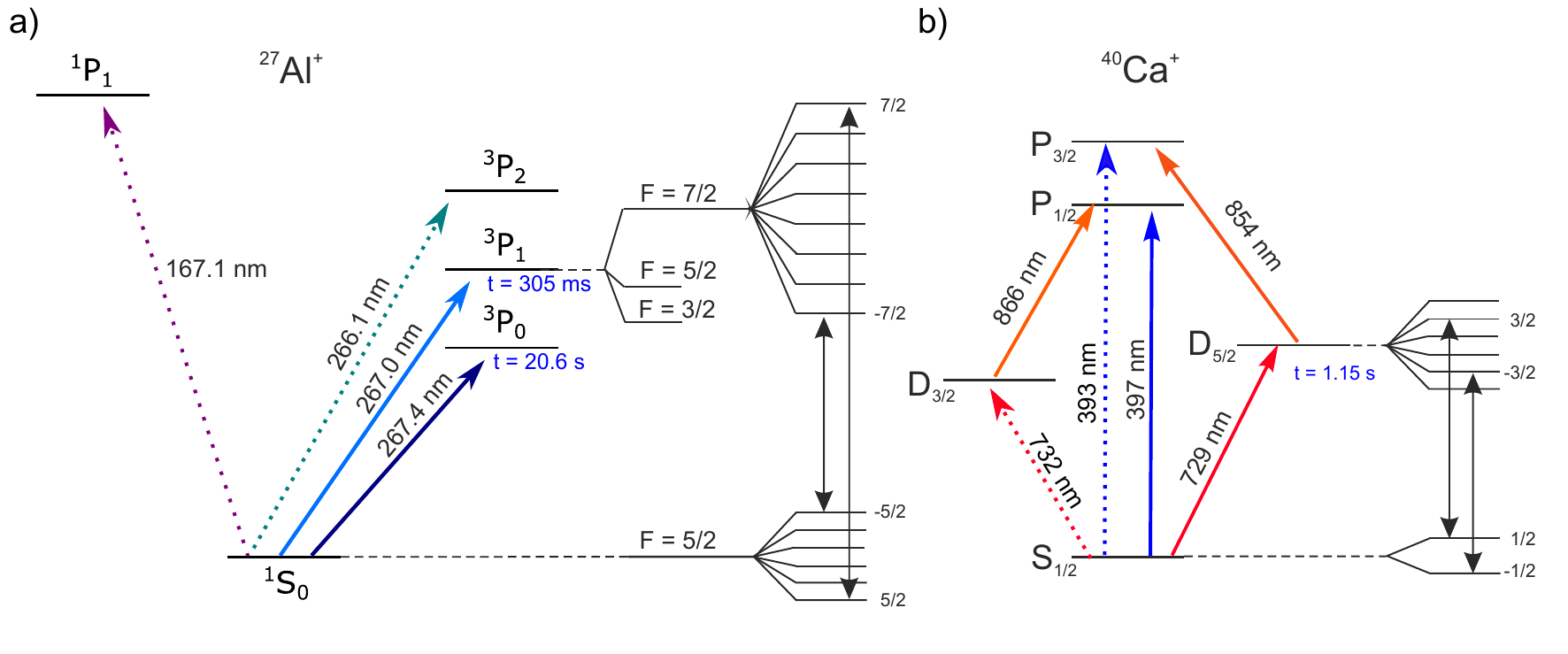}
 \caption{\label{fig:levels}
 Lowest energy levels of the $^{27}\mathrm{Al}^+$ and $^{40}\mathrm{Ca}^+$ . (a) $^{27}\mathrm{Al}^+$: There are three transitions to the triplet P
states, namely the hyperfine-induced clock transition at 267.4\,nm, the intercombination transition at 267.0\,nm and the magnetic quadrupole transition at 266.1\,nm. Hyperfine interaction splits the $^3$P$_1$ level into three components. Excitation of the intercombination line on the $^1\mathrm{S}_0\leftrightarrow ^3\mathrm{P}_1, F=7/2$ transition enables one to prepare the ion in a single Zeeman ground state and to detect whether the ion was excited to the $^3\mathrm{P}_0$ state by a laser pulse probing the clock transition. In a magnetic field that lifts the degeneracy of the Zeeman states, the frequency of the intercombination line can be determined by probing the two stretched transitions indicated in the plot by black arrows. (b) $^{40}\mathrm{Ca}^+$: The $\mathrm{S}_{1/2}\leftrightarrow \mathrm{D}_{5/2}$ electric quadrupole transition at 729~nm constitutes a frequency standard that can be interrogated by exciting two or more of the ten available Zeeman transition. All transitions denoted by solid lines can be probed by lasers that are available in our experimental setup. For further details, see main text.

}
\end{center}
\end{figure}

From this perspective, the $^{1}\mathrm{S}_{0}\leftrightarrow\,^{3}\mathrm{P}_{0}$ clock transition in Al$^+$ (see Fig.~\ref{fig:levels}) has very attractive features, combining an extremely narrow linewidth of about 8\,mHz at 267.4~nm, which gives rise to a high \textit{Q} factor of $\nu/\Delta\nu\approx10^{17}$, with high immunity to perturbations by external fields. The transition has a low dc second-order Zeeman shift, a very small black-body radiation shift, and a negligibly small electric quadrupole shift. The total contribution of this shifts to the total systematic relative uncertainty of the clock has been reported with values below $\Delta\nu/\nu\approx 8 \times 10^{-18}$ \cite{Rosenband:2008, Rosenband:2006,Beloy:2017} . For this reason, there are ongoing efforts in several laboratories for constructing an optical frequency standard based on $^{27}$Al$^+$ \cite{Schmidt:2005,Wuebbena:2014,Xu:2016,Cui:2018}. In addition, singly charged aluminium ions are a promising candidate for the construction of a transportable optical clock with low fractional systematic uncertainties \cite{Hannig:2019}.

Experiments with Al$^+$ are complicated by the fact that the dipole-allowed transition $^{1}\mathrm{S}_{0}\leftrightarrow\,^{1}\mathrm{P}_{1}$ that could be used for detecting the quantum state of the ion via state-dependent fluorescence measurements and for preparing the ion in a desired initial state has a wavelength of 167~nm. This makes the operation of Al$^+$ clocks technically more challenging than the one of other ion clocks. Moreover, its linewidth of about 220~MHz is rather large so that Doppler cooling on this transition would lead to minimum temperatures that are a magnitude above the ones reached when Doppler cooling Ca$^{+}$, Sr$^{+}$, Hg$^{+}$ or Yb$^{+}$ ions \cite{kramida2018nist}. The problem of detecting the quantum state of the aluminiun ion is overcome by quantum logic spectroscopy \cite{Wineland:2002}, a technique that enables detecting the quantum state of a trapped atomic or molecular ion \cite{Schmidt:2005,Rosenband:2007,Hempel:2013,Wan:2014,Wolf:2016,Chou:2017} by transferring the information to a co-trapped ion that can be easily read out. 

Excitation of Al$^+$ ions on their $^{1}\mathrm{S}_{0}\leftrightarrow\,^{3}\mathrm{P}_{1}$ intercombination line at 267~nm by coherent laser pulses is a technique that enables a set of useful tools for the construction of an optical frequency standard. As the $^{3}\mathrm{P}_{1}$ state has a lifetime of about $\sim 300\,\mu$s \cite{Traebert:1999}, the intercombination line can be used for initializing the ion in a pure electronic state by optical pumping, and, even more importantly, for high-fidelity state detection after probing its clock transition by repetitive quantum non-demolition measurements based on quantum logic spectroscopy \cite{Hume:2007}. Despite the importance of this transition, to our knowledge, a measurement of its transition frequency has only been carried out once before and reported in a conference abstract \cite{Rosenband:2005}.

In the present paper, we describe quantum logic spectroscopy experiments with a two-ion crystal composed of a $^{27}$Al$^+$ ion and a $^{40}$Ca$^+$ ion. We use $^{40}$Ca$^{+}$ as the logic ion primarily because of its similar mass and a relatively narrow cooling transition, which enables efficient sympathetic cooling~\cite{Wuebbena:2012}. Moreover, the $\mbox{S}_{1/2}\leftrightarrow \mbox{D}_{5/2}$ quadrupole transition can be used for a direct clock comparison. Using a frequency comb, we measure the transition frequency of the   $^{1}\mathrm{S}_{0}\leftrightarrow\,^{3}\mathrm{P}_{1}$ line in $^{27}$Al$^+$ by interleaving probing of this transition with probing of the $\mbox{S}_{1/2}\leftrightarrow \mbox{D}_{5/2}$ quadrupole transition in $^{40}$Ca$^+$, which is used as a frequency reference. As a test for systematic errors in the  measurement of our $^{40}$Ca$^+$ frequency reference, its transition frequency is compared to the ones measured in two other $^{40}$Ca$^+$ experiments, which are located in neighbouring laboratories. In addition, we also determine the \textit{g}-factor of the $^{3}\mathrm{P}_{1}, F=7/2$ level.

The paper is organised as follows: the experimental apparatus is described in section~\ref{apparatus}. In section~\ref{results}, we describe the preparation of the mixed $^{40}$Ca$^{+}$--$^{27}$Al$^{+}$ ion crystal, the quantum logic spectroscopy protocol, and relative frequency measurements of the quadrupole transition in $^{40}$Ca$^+$. In section~\ref{section:frequency}, we describe the absolute frequency measurement of the $^{1}\mathrm{S}_{0}\leftrightarrow\,^{3}\mathrm{P}_{1}$ transition, including a discussion of relevant systematic frequency shifts. We conclude the paper by a brief discussion of the results and preliminary measurements of the $^{1}\mathrm{S}_{0}\leftrightarrow\,^{3}\mathrm{P}_{0}$ clock transition.
 
\section{Experimental apparatus} \label{apparatus}

The experimental apparatus that we designed for quantum-logic experiments with Ca$^{+}$ and Al$^{+}$ ions has been previously described in detail in the Ref.~\cite{Guggemos:2015}. In the following, we provide a brief summary of its main features.
Ions are stored in a blade-style linear ion trap mounted in an ultra-high vacuum chamber at a residual gas pressure of roughly 10$^{-11}$\,mbar. 
A magnetic field of about 4\,G, which defines the quantisation axis along the symmetry axis of the trap, is created by a pair of coils driven by a current source with a long-term fractional current stability of about $|\Delta I/I|\sim 3\cdot 10^{-6}$. Singly charged Al and Ca ions are created via two-step photoionisation from neutral atoms obtained by short-pulse ablation from two different targets. 
For Doppler cooling and detecting the Ca$^{+}$ ions, we use the light of a frequency-doubled extended-cavity diode laser at 397\,nm resonant with the $\mathrm{S}_{1/2}\leftrightarrow \mathrm{P}_{1/2}$ transition (see Fig.~\ref{fig:levels}). Two extended-cavity diode lasers, one emitting at 866~nm and the other at 854\,nm are used as repumpers. The fluorescence emitted by calcium ions at 397~nm is recorded by a photomultiplier tube and an electron-multiplying CCD camera.

\begin{figure}[tb]
\begin{center}
\includegraphics[width=0.8\textwidth]{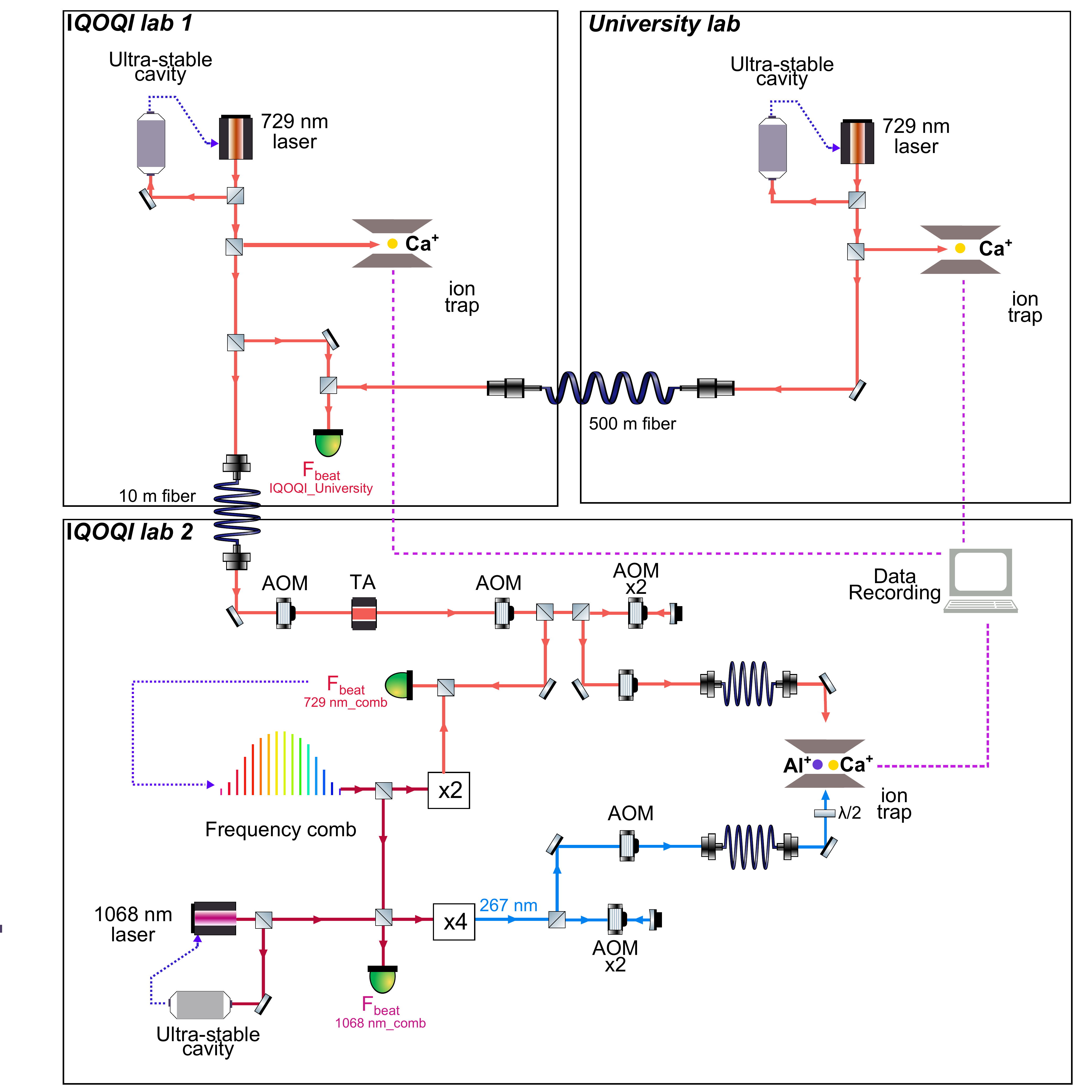}
\caption{\label{fig:comb}
Overview of the optical setup used for probing and measuring the narrow transitions in $^{40}$Ca$^+$ and $^{27}$Al$^+$ in \textit{IQOQI lab2}. For interrogation of the $^{27}$Al$^+$ transition, a diode laser at 1068\,nm is stabilised to a high-finesse cavity and frequency-quadrupled to 267\,nm. The frequency of the 1068\,nm laser is constantly measured by a frequency comb whose repetition rate is locked to the frequency of the ultra-stable laser at 729\,nm located in \textit{IQOQI lab1}. A fraction of this laser's output is amplified in a tapered amplifier (TA) and delivered to the ion trap in \textit{IQOQI lab2}. An optical fibre linking the \textit{IQOQI labs} to the \textit{University lab}, which houses an independent frequency-stable Ti:Sa laser at 729~nm, enables a performance evaluation of the two 729~nm laser systems and probing of systematic frequency shifts of the quadrupole transition in Ca$^+$ by a simultaneous interrogation of Ca$^+$ ions in three different ion traps. To keep the diagram simple, we sketch only \textit{IQOQI lab2}'s acousto-optic modulators (AOMs), which are used for frequency-shifting and intensity-switching the laser beams.

}
\end{center}
\end{figure}

Figure~\ref{fig:comb} shows the laser setup used for probing and measuring the narrow transitions in $\mathrm{Ca}^+$ and $\mathrm{Al}^+$. For coherent manipulation of the electronic and motional state of the calcium ion, the $\mathrm{S}_{1/2}\leftrightarrow \mathrm{D}_{5/2}$ quadrupole transition at 729\,nm is excited by a titanium-sapphire laser locked to a high-finesse cavity yielding a
spectral linewidth on the order of 1\,Hz on the time scale of a few seconds; the fibre-induced noise between the laser and the cavity is suppressed by using a similar arrangement as in Ref.~\cite{Ma:1994}. The transition frequency can simultaneously be compared with frequency measurements obtained in two other $^{40}$Ca$^+$ experiments in nearby laboratories, designated from now on in the paper as \textit{IQOQI lab 1} and \textit{University lab} (see Fig.~\ref{fig:comb}). The experiment setups of both laboratories are described in Refs.\cite{Hempel:2013,Chwalla:2009}, respectively. The fibres connecting different laboratories have optical path lengths that are interferometrically stabilised. 

For excitation of the $(3s^{2})\,^{1}\mathrm{S}_{0} \leftrightarrow \, (3s3p)\,^{3}\mathrm{P}_{1}, F=7/2$ transition in $^{27}$Al$^{+}$ a frequency-quadrupled extended-cavity diode laser is used, with a maximum output power of 20\,mW at 267.0\,nm. The fundamental radiation at 1068\,nm of this laser system is stabilised to a high finesse cavity ($\mathcal{F}= 300 000$). Additionally, a frequency-quadrupled fibre laser is available for excitation of the clock transition at 267.4\,nm. Beat measurements between the two lasers (tuned to the same wavelength) reveal a linewidth of 1.6\,Hz FWHM or less at 4\,s at the fundamental frequency of the lasers. In order to guide the 267.0\,nm light to the ion, we implemented a solid--core photonic crystal fibre made from fused silica~\cite{Colombe:2014}. This arrangement has the advantage of providing a clean beam profile, thus minimizing the amount of UV light that could illuminate the trap electrodes giving rise to the emission of undesired photo-electrons. The Al$^{+}$/Ca$^{+}$ two-ion crystal is illuminated by approximately 200\,$\mu$W of 267.0\,m light by a laser beam parallel to the weak axis of the trap with a beam diameter of $\sim 65\,\mu$m. 
During the quantum logic spectroscopy experiments the frequency of the 1068\,nm laser is measured with a femtosecond frequency comb whose repetition rate is locked to the frequency of the ultra-stable laser at 729\,nm located at \textit{IQOQI lab1}. 

\section{Experimental results} \label{results}
Our experiments start by loading a single Ca$^+$ ion into the trap  and optimizing its cooling and fluorescence measurements. Typical oscillation frequencies of a single Ca$^+$ ion are 820~kHz along the weak axial direction of the trap and about 1.5~MHz along the radial directions. Micromotion caused by residual electric stray fields is detected by probing the micromotion sidebands of the $\mathrm{S}_{1/2}$ to $\mathrm{D}_{5/2}$ transition. It is compensated by applying additional voltages to two compensation electrodes shifting the ion in the radial plane of the linear trap and by unbalancing the dc-voltages applied to the tip electrodes providing the axial confinement. Modulation indices of less than $10^{-2}$ on the transition used for micromotion compensation are obtained at the end of the procedure.
\subsection{Trapping, cooling and motional analysis of two-ion Ca$^+$/Al$^+$ crystals } \label{section:mixedcrystals}
As described in Ref.~\cite{Guggemos:2015}, a $\mathrm{Ca}^{+}\!/\mathrm{Al}^{+}$ ion crystal is created by loading an Al$^+$ via laser ablation. In experiments probing the intercombination line, mixed ion crystal lifetimes of several hours are typically observed, limited by background gas collisions leading to ion loss or chemical reactions giving rise to conversion of atomic ions to unwanted molecular ions such as AlH$^{+}$. The mass of molecular ions created in the trap is inferred from ``tickling'' measurements, in which the fluorescence of the calcium ion is measured as a function of the frequency of a weak electric field capable of exciting the in-phase collective motion of the crystal along its axis.

\begin{figure}[tb]
\begin{center}
\includegraphics[width=.8\textwidth]{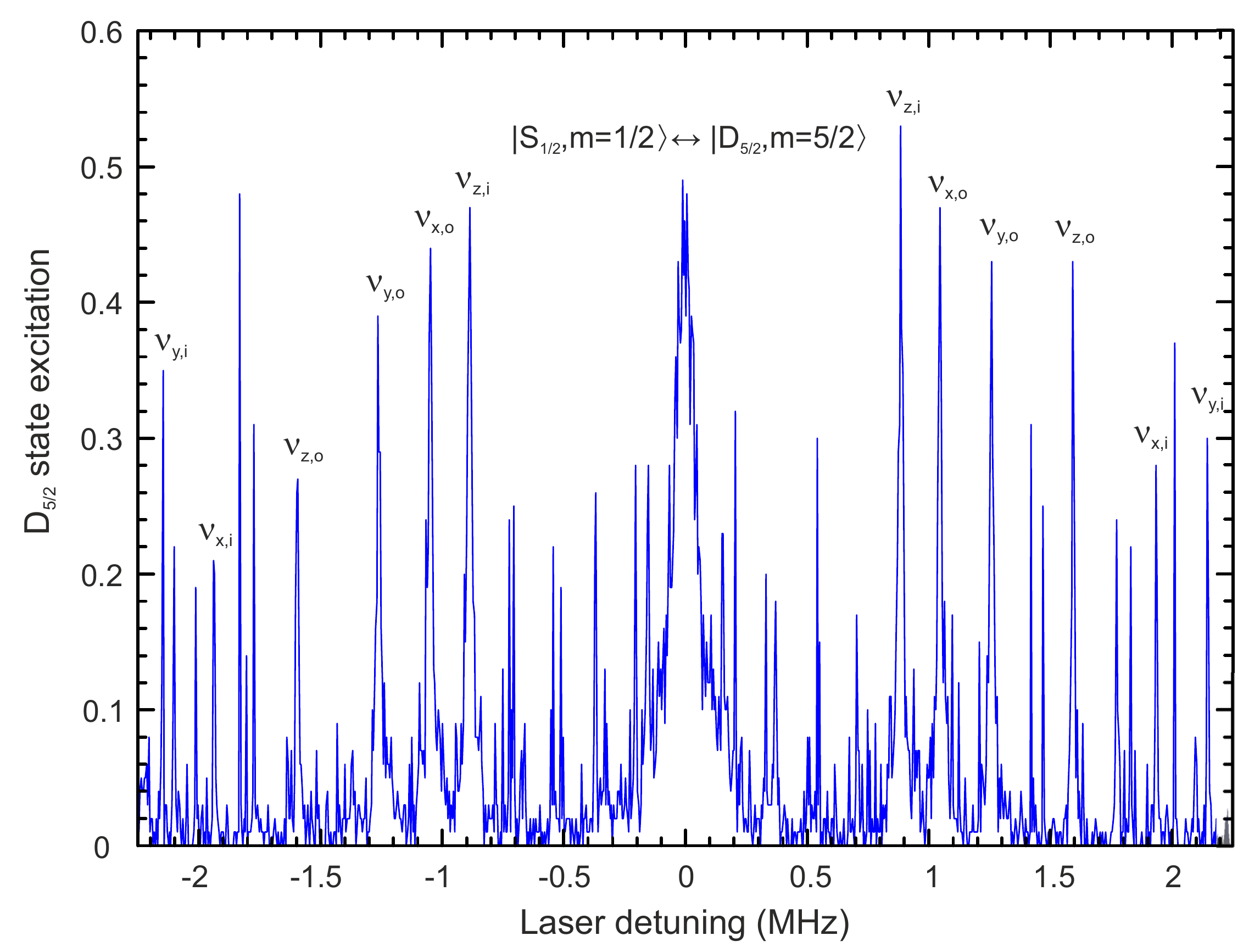}

\caption{\label{fig:spectrum}
Excitation spectrum of a calcium ion probed at 729~nm in a mixed Al$^{+}$/Ca$^{+}$ ion crystal at typical trapping conditions. The center peak represents the Ca$^{+}\,S_{1/2},m_{J}=1/2\leftrightarrow\,D_{5/2},m_{J}=5/2$ carrier transition with its red and blue motional sidebands of the mixed crystal symmetrically placed to the left and right, respectively. The axial in- and out-of-phase modes are visible at $\nu_{z,i}=888$\,kHz and $\nu_{z,0}=1.596$\,MHz respectively. The degeneracy of the radial potential is lifted by a dc-voltage applied to a pair of blade electrodes so that the radial modes are observed at approximately $\nu_{y,o}=0.9\,$MHz, $\nu_{x,o}=1.1\,$MHz, $\nu_{y,i}=1.9\,$MHz, and $\nu_{x,i}=2.1\,$MHz. Other peaks appearing correspond to second-order sidebands.
}
\end{center}
\vspace{-7mm}
\end{figure}

A second tool for investigating the collective motion of the two-ion crystal is provided by measurements of absorption sideband spectra on the quadrupole transition in Ca$^{+}$. Figure~\ref{fig:spectrum} shows a part of an absorption spectrum obtained by a laser beam with overlap with all six collective modes of motion, which give rise to upper and lower vibrational sidebands of the carrier transition shown in the center of the plot. 

For carrying out the quantum logic spectroscopy protocol described further below (section.~\ref{section:QLS}), one common vibrational mode of the crystal has to be cooled to its ground state. We observe heating rates of the axial in-phase (z$_{in}$) and out-of-phase (z$_{out}$) mode of $\approx$ 70\,phonons/s and $\approx$ 0.8\,phonons/s, respectively. Due to its lower heating rate, the z$_{out}$ mode was chosen as the bus mode for the quantum state transfer. Possible dispersive frequency shifts by cross-mode coupling to other collective modes \cite{Roos:2008a} are at the level of a few Hertz and thus too small to be relevant for the experiments described in this paper. In our experiments, we cool both axial vibrational modes to mean phonon numbers below 0.05 by sideband cooling the calcium ion, in order to minimise coupling strength fluctuations when resonantly exciting sideband transitions in Al$^+$ or Ca$^+$.

\subsection{Preparation of the Al$^+$ ion in a specific Zeeman ground state}\label{Sec:Alzeemanstate}

The Al$^+$ ion is initialised in a pure electronic state by optical pumping on the intercombination line. In our current experiment, there is only a single laser beam available for exciting the ion from a direction that is parallel to the quantisation axis set by the magnetic field. By making its polarisation circular, it is used to populate either one of the $m=\pm 5/2$ stretched Zeeman states using a series of carrier $\pi$-pulses exciting the five $m\leftrightarrow m\pm 1$ Zeeman transitions that can be used for pumping out the undesired Zeeman states.

We also tested whether frequency-resolved optical pumping to the stretched states could be achieved by making the beam's polarisation linear. This approach comes with the prospect of being able to optically pump into any Zeeman state or to rapidly switch between populating one or the other stretched state by optical pumping without resetting the beam's polarisation. This approach enabled us to excite all available $\sigma_+ (\sigma_-)$ Zeeman transitions using the protocol described below. We found, however, that the pumping worked less reliably as compared to the one by a circularly polarised beam. We attribute these difficulties to coherent population trapping and ac-Stark shifts by the unwanted polarisation component as adjacent Zeeman ground states have energy shifts of less than 10~kHz at a magnetic field of 4~G. From this perspective, it seems more promising to add fast polarisation control or a second laser beam with a different polarisation to the experimental setup.


\subsection{Quantum logic spectroscopy of the $^{1}\mathrm{S}_{0},F=5/2 \,\leftrightarrow \,^{3}\mathrm{P}_{1}, F=7/2$ transition} \label{section:QLS}

Quantum logic spectroscopy experiments probing the intercombination line in Al$^+$ are carried out by using the following steps for quantum state initialisation, excitation, and probing:
\begin{enumerate}

\item {\sl Logic ion initialisation:} The ion crystal is prepared in the Lamb-Dicke regime by 5~ms of Doppler cooling on the $\mathrm{S}_{1/2}\leftrightarrow\mathrm{P}_{1/2}$ transition of Ca$^+$ at 397\,nm. Light at 866\,nm and 854\,nm is used to optically pump out the metastable $\mathrm{D}_{3/2}$ and $\mathrm{D}_{5/2}$ states of the logic ion. A circularly polarised beam at~397 nm oriented in the direction of the quantisation axis initialises the calcium ion in one of the $\mathrm{S}_{1/2}$ ground states.

\item {\sl Spectroscopy ion initialisation:} The Al$^+$ ion is optically pumped into either one of the two stretched $m_F=\pm 5/2$ Zeeman states of the $^{1}\mathrm{S}_{0}$ ground state by exciting the $^{1}\mathrm{S}_{0},F=5/2$ to $^{3}\mathrm{P}_{1}, F=7/2$ intercombination line with a circularly polarised beam at 267~nm counterpropagating to the circularly polarised beam at 397~nm. A series of laser pulses driving the five $m_F\leftrightarrow m_{F^\prime}\pm 1$ Zeeman transitions ($-5/2\le m_F<5/2$) with pulse areas of $\pi$ followed by a waiting time of 300~$\mu$s pushes the population towards the target state. This sequence of pulses is repeated 10 times in order to populate the stretched Zeeman state.

\item {\sl Ground-state cooling and logic ion preparation:} Both axial modes are sideband-cooled close to the ground state by two laser pulses on the quadrupole transition at 729~nm with a total duration of 7~ms. To minimise the average vibration quantum number of the out-of-phase mode used for the quantum logic state transfer, the in-phase-mode is cooled prior to cooling the out-of-phase mode. Finally, a carrier $\pi$-pulse transfers the Ca$^+$ ion into the $D_{5/2}$ state. 

\item {\sl Spectroscopy ion probing:} The $|^{1}\mathrm{S}_{0},F=5/2, m_{F}=\pm5/2\rangle \leftrightarrow \,|^{3}\mathrm{P}_{1}, F=7/2 ,m_{F^\prime}=\pm7/2\rangle$ carrier transition is probed by either a single pulse or in a Ramsey experiment by a pair of $\pi/2$ pulses separated by a waiting time. 

\item {\sl Quantum state transfer:} For state mapping to the logic ion, a $\pi$-pulse of the upper motional sideband of the out-of-phase mode on the $|^{1}\mathrm{S}_{0},F=5/2, m_{F}=\pm5/2\rangle \leftrightarrow \,|^{3}\mathrm{P}_{1}, F=7/2 ,m_{F^\prime}=\pm7/2\rangle$ transition adds a phonon to the system if the previous carrier excitation was unsuccessful. The phonon is mapped to a change of the electronic state in Ca$^+$ by a consecutive $\pi$-pulse on the upper motional sideband of the same mode on the $\mathrm{S}_{1/2}\leftrightarrow\,\mathrm{D}_{5/2}$ transition, which returns the calcium ion to its electronic ground state. As a result, the calcium ion is found in $D_{5/2}$ if the aluminium ion was excited to the metastable state by the carrier excitation pulse of step (iv).

\item {\sl Quantum state detection:} The quantum state of Ca$^+$ is measured via electron shelving.    
  
\end{enumerate}
		
\begin{figure}[tb]
\begin{center}

\includegraphics[width=1\textwidth]{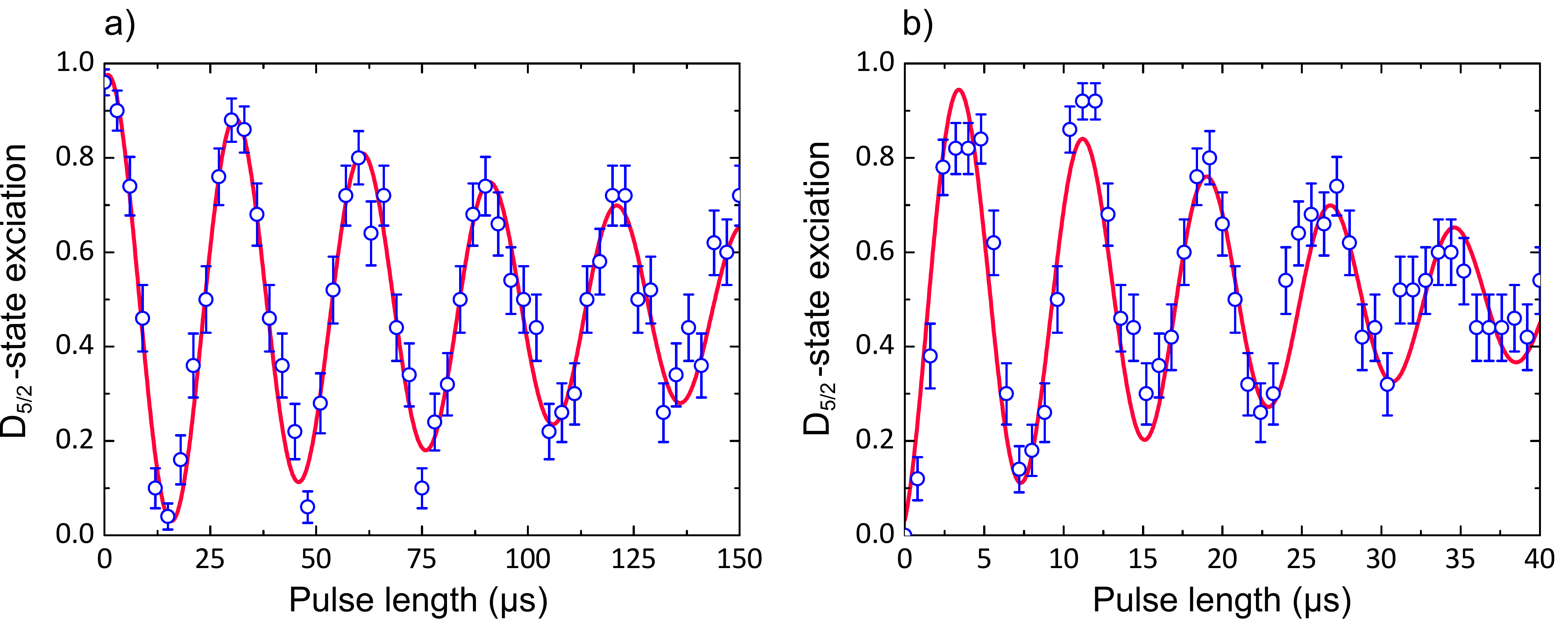}

\caption{\label{fig:rabi}
(a) Rabi oscillations on the blue sideband (z$_{out}$) of the $(3s^{2})\,^{1}\mathrm{S}_{0},m_{F}=-5/2 \leftrightarrow \, (3s3p)\,^{3}\mathrm{P}_{1}, F=7/2,m_{F}=-7/2$ transition detected by quantum logic spectroscopy. After $15\,\mu$s approximately 95\% of the population is transferred to the excited state. The imperfections of this $\pi$-pulse give rise to the baseline in the Ramsey patterns. The fidelity is limited by high frequency noise of the laser and temperature fluctuations of the axial in-phase mode. The error bars are determined from quantum projection noise. The red curve, an exponentially decaying sine function including a constant offset, is fitted to the experimental data. (b) Rabi oscillations on the carrier transition for identical parameters. The signal is obtained by adding a carrier excitation pulse of variable length to the pulse sequence prior to the blue-sideband $\pi$-pulse.   
}
\end{center}
\end{figure}

\noindent Figure~\ref{fig:rabi} shows an application of this protocol that was used for recording Rabi oscillations on the blue sideband and the carrier transition of the $|^{1}\mathrm{S}_{0},F=5/2, m_{F}=5/2\rangle \leftrightarrow |\,^{3}\mathrm{P}_{1}, F=7/2 ,m_{F^\prime}=7/2\rangle$ transition. Note, that in this case, we omit step (iv) and vary the duration fo the sideband pulse in step (v). It demonstrates that $\pi$-pulses can be realised on the sideband with a fidelity of about 95~\% and that the Al$^+$ ion can be coherently excited to the metastable state by a carrier $\pi$-pulse with a duration of about $4\,\mu$s. On the carrier transition, the loss of contrast versus time is predominantly caused by high-frequency phase noise of the quadrupled diode laser system. This assumption is corroborated by the fact that the ratio $\gamma/\Omega$ between the damping rate $\gamma$ of the oscillations and the carrier Rabi frequency $\Omega$ is not constant but increases with $\Omega$ in qualitative agreement with the measured spectral phase noise of the laser. On the sideband, the contrast loss is due to laser phase noise and loss of population from the upper state by spontaneous emission.

\begin{figure}[tb]
\begin{center}
\includegraphics[width=1.0\textwidth]{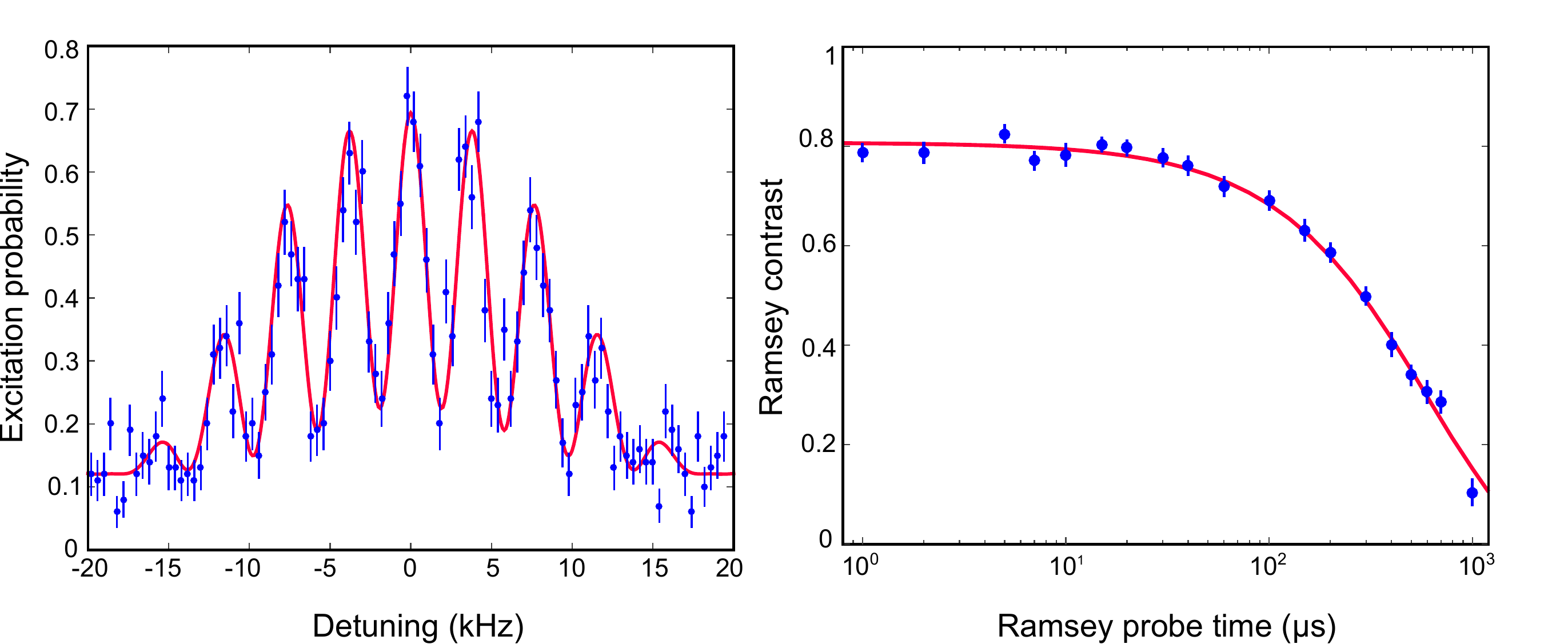}
\caption{
\label{fig:Ramsey}
(a) Ramsey experiment on the $^{27}$Al$^{+}$ $(3s^{2})\,^{1}\mathrm{S}_{0},m_{F}=-5/2\leftrightarrow\, (3s3p)\,^{3}\mathrm{P}_{1}, F=7/2,m_{F^{\prime}}=-7/2$ carrier transition: measured excitation as a function of laser detuning (blue points)  for Ramsey pulses of 50\,$\mu$s duration and a waiting time of 200\,$\mu$s. The data is fitted  by a numerical integration of the master equation describing the Ramsey experiment (solid line), including spontaneous decay of the $^{3}\mathrm{P}_{1}$ level. The resulting excitation probability is adjusted for a frequency offset, an excitation baseline and a reduction in contrast. (b) Ramsey contrast versus waiting time on the same transition. Fitting an exponential decay curve to the data yields a contrast of 0.81(1) for very short Ramsey times and a decay time of $\tau=598(20)\,\mu$s that agrees within its error bar with twice the lifetime of the $^{3}\mathrm{P}_{1}$ level. 
}
\end{center}
\end{figure}

For a further investigation of the coherence time on the intercombination line, Ramsey experiments on the carrier transition were carried out using the quantum logic spectroscopy protocol for state read-out.  Figure~\ref{fig:Ramsey}(a) shows the Ramsey signal as a function of laser detuning, whereas Fig.~\ref{fig:Ramsey}(b) shows the Ramsey contrast as a function of the waiting time between the $\pi/2$-pulses. At short waiting times a Ramsey contrast of 0.81(1) is obtained which is limited by imperfect $\pi/2$ and state mapping pulses. At longer waiting times, the contrast decays exponentially with time, in agreement with the expected contrast loss caused by spontaneous decay of the $^3\mathrm{P}_1$ state.

\subsection{Probing the $\mathrm{S}_{1/2}\leftrightarrow\mathrm{D}_{5/2}$ transition in $^{40}$Ca$^+$\label{section:ProbingCa}}

The $\mathrm{S}_{1/2}\leftrightarrow\,\mathrm{D}_{5/2}$ transition in $^{40}$Ca$^+$ was used as a stable and calibrated frequency reference \cite{Huang:2016, Chwalla:2009, Matsubara:2012} for the 729~nm laser and as a magnetic field sensor. In order to keep track of any frequency drift $\Delta f$ of the 729\,nm laser or any magnetic field fluctuations $\Delta B$ we probed two different Zeeman transitions, $|\mathrm{S}_{1/2},m=\pm 1/2\rangle\leftrightarrow\,|\mathrm{D}_{5/2}, m^\prime=\pm 3/2\rangle$, as illustrated in Fig.~\ref{fig:levels}, with Ramsey experiments composed of $\pi/2$-pulses of $50\,\mu$s duration that were separated by a 2~ms waiting time. The frequency shift of the laser $\Delta$f is given by the arithmetic mean of the measured detunings from the two transitions. As the $g$-factors of the electronic states are well-known, the frequency splitting of the Zeeman transitions provides also a measure of the absolute value of the dc-magnetic field fluctuations $\Delta$B. Choosing two opposite Zeeman transitions has the advantage of cancelling not only the first-order dc-Zeeman shift but also possible ac-Zeeman shifts caused by currents oscillating at the trap drive frequency.

\begin{figure}[tb]
\begin{center}
\includegraphics[width=1\textwidth]{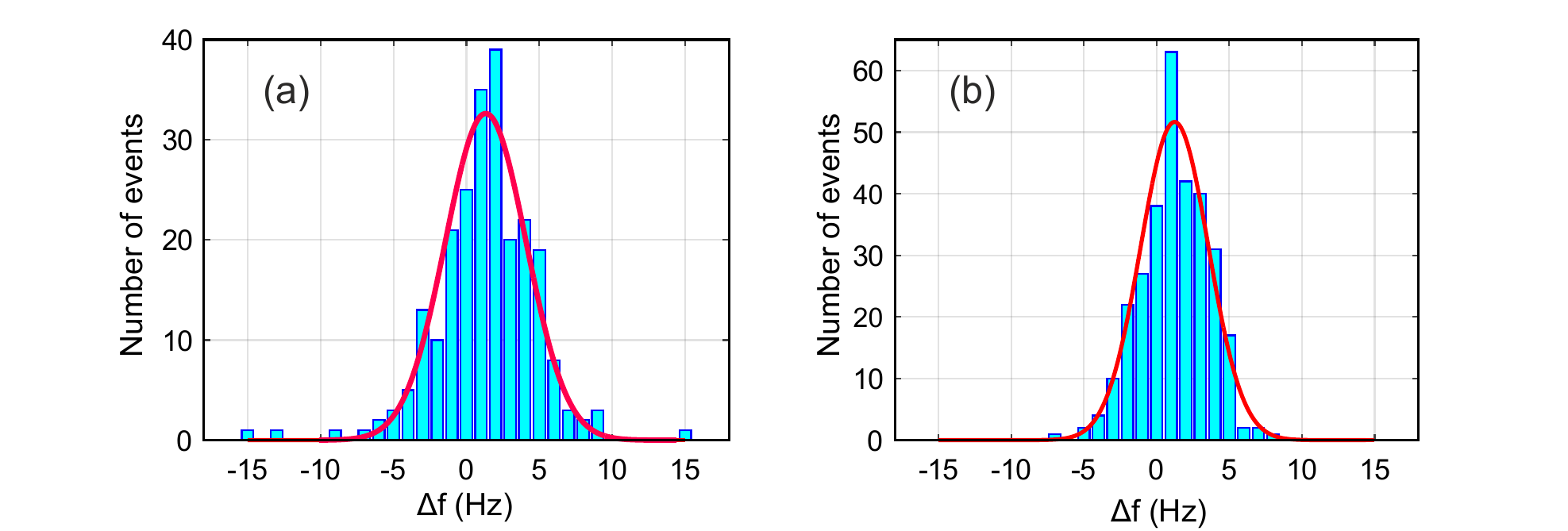}
\caption{Comparison of the measured frequency differences of the $\mathrm{S}_{1/2}\leftrightarrow\,\mathrm{D}_{5/2}$ transition of a single $\mathrm{Ca}^+$ ion in the present setup (\textit{IQOQI lab~2}) with two similar experimental setups, (a) \textit{IQOQI lab~1}, and (b) \textit{University lab}. For this histogram, the data was binned into one-minute time intervals. For a detailed description of the measurement scheme, see section~\ref{section:ProbingCa}. A Gaussian fit of the histogram distribution (red line) in both cases yields line widths of 2.77(0.15)~Hz and 2.33(0.11)~Hz, respectively, which are consistent with measurement uncertainties caused by quantum projection noise.
\label{fig:LabComparison}
}
\end{center}
\end{figure}

To ensure that systematic frequency shifts do not cause any significant error in the determination of the frequency of the intercombination line in Al$^+$, a direct comparison of the calcium transition frequency measured in three different laboratories was made. The frequency measured in our experimental setup (\textit{IQOQI lab~2}) was compared to the frequency measured in two similar experimental setups (see Fig.~\ref{fig:comb}). The setups of \textit{IQOQI lab1} and \textit{IQOQI lab 2} share the same 729~nm laser while the \textit{University lab} has its own independent laser setup. Laser beat measurements show, after subtraction of a linear frequency drift, that the two lasers differ by no more than 10~Hz over the course of several hours. For the frequency comparison, two Zeeman transitions, $\mathrm{S}_{1/2},m=\pm 1/2\leftrightarrow\,\mathrm{D}_{5/2}, m^\prime=\pm 3/2$, were probed with Ramsey experiments of 3~ms duration during a period of five hours. The data was subsequently binned into one-minute intervals and the frequency difference between two setups was evaluated. Fig.~\ref{fig:LabComparison} shows histograms of the measured frequency differences with the other two setups. 

The frequency measured in the \textit{IQOQI lab~2} experimental setup differed by 1.3(2)~Hz from the value measured in \textit{IQOQI lab~1}, and by 1.2(2)~Hz from the value found in \textit{University lab}. We attribute the offsets between the labs mainly to different electric quadrupole shifts due to different axial trapping confinements and stray electric field gradients, and to a lesser degree to AC Stark shifts caused by stray light. A detailed investigation of the systematic frequency shift of a single Ca$^{+}$ ion clock can be found in Ref.~\cite{Chwalla:2009}. After subtraction of known systematic shifts, the frequencies agree with each other to better than 1~Hz.

\section{Frequency measurement of the intercombination transition} \label{section:frequency}
In the following, we will describe how we infer the absolute frequency of the intercombination line from the measurement results we obtained when probing the transition and discuss the corresponding error budget.

\subsection{Measurement results}

For a measurement of the $(3s^{2})\,^{1}S_{0}\leftrightarrow \,(3s3p)\,^{3}P_{1}, F=7/2$ transition, we probe the Al$^+$ ion with Ramsey experiments as described in subsection \ref{section:QLS}. The fundamental frequency of the laser probing the transition is recorded with the frequency comb whose repetition rate is locked to the frequency of the ultrastable laser at 729~nm. The frequency of the latter is monitored by Ramsey experiments on Ca$^+$ that are interleaved with experiments probing the Al$^+$ ion. In this way, the frequency of the intercombination line can be determined with respect to the well-known $\mathrm{S}_{1/2}\leftrightarrow\,\mathrm{D}_{5/2}$ transition frequency of $^{40}\mathrm{Ca}^+$. The comb measurements show that frequency fluctuations of the light probing the intercombination line in Al$^+$ are typically about 60~Hz over the course of fifteen minutes. 

We interrogated the $|^{1}\mathrm{S}_{0}, F=5/2,m_F=5/2\rangle\leftrightarrow\,|^{3}\mathrm{P}_{1}, F=7/2,m_{F^\prime}=7/2\rangle$ Zeeman transition with Ramsey experiments composed of $\pi/2$-pulses of $50\,\mu$s duration separated by a waiting time of either 100 or 200~$\mu$s. The phase $\phi$ of the second $\pi/2$-pulse was shifted with respect to the first one by either $0$, $\pm\pi/2$, or $\pi$, in order to extract both the laser-ion detuning and the contrast of the Ramsey fringe. For each setting of $\phi$, the experiment was repeated 100 times. Next, another four sets of 100 experiments were carried out probing two Zeeman $\mathrm{S}_{1/2}\leftrightarrow\,\mathrm{D}_{5/2}$ transitions in $^{40}\mathrm{Ca}^+$ as described in subsection~\ref{section:ProbingCa}, with the phase of the second Ramsey pulse shifted by $\phi=\pm\pi/2$, for a measurement of the detuning of the laser at 729~nm from the atomic transition and the magnetic bias field. This cycle of eight different experiments was repeated fifty times followed by a recalibration of experimental parameters. The same type of experiment was also carried out for a measurement of the $|^{1}\mathrm{S}_{0}, F=5/2,m_F=-5/2\rangle\leftrightarrow\,|^{3}\mathrm{P}_{1}, F=7/2,m_{F^\prime}=-7/2\rangle$ Zeeman transition. The absolute value of the transition frequency is obtained with respect to the transition frequency of the $S_{1/2}\leftrightarrow D_{5/2}$ transition in $^{40}\mathrm{Ca}^+$, $\nu_{\mathrm{Ca}}=411\,042\,129\,776\,398(4)$~Hz, where we took the weighted average of the frequency measurements reported by three different groups in Refs. \cite{Huang:2016, Chwalla:2009, Matsubara:2012}.

Our evaluation of the $^{1}\mathrm{S}_{0}, F=5/2\leftrightarrow\,^{3}\mathrm{P}_{1}, F=7/2$ transition frequency is based on a total of 18 sets of measurements. For each set, the uncertainty of the measured average deviation of the laser frequency from the atomic transition is predominantly given by quantum projection noise. 
For the probed Zeeman transitions, the transition frequency $f$ at non-zero magnetic field $B$ is related to the transition frequency $f_0$ at zero magnetic field by 
$f=f_0+\frac{\mu_B}{h}({\textstyle \frac{7}{2}}g_{^{\,3\!}P_1,F^\prime=7/2}-{\textstyle\frac{5}{2}}g_{^{\,1\!}S_0})s_\pm B$ where $s_\pm=\pm 1$ depending on which of the two stretched state transitions is probed, with $\mu_B$ the Bohr magneton and $h$ Planck's constant. For each measurement set, we determine the magnetic field strength by measuring the frequency splitting of the two Zeeman transitions probed in Ca$^+$ with an uncertainty of about 1~$\mu$Gauss.

\begin{figure}[tb]
\begin{center}
\includegraphics[width=0.8\textwidth]{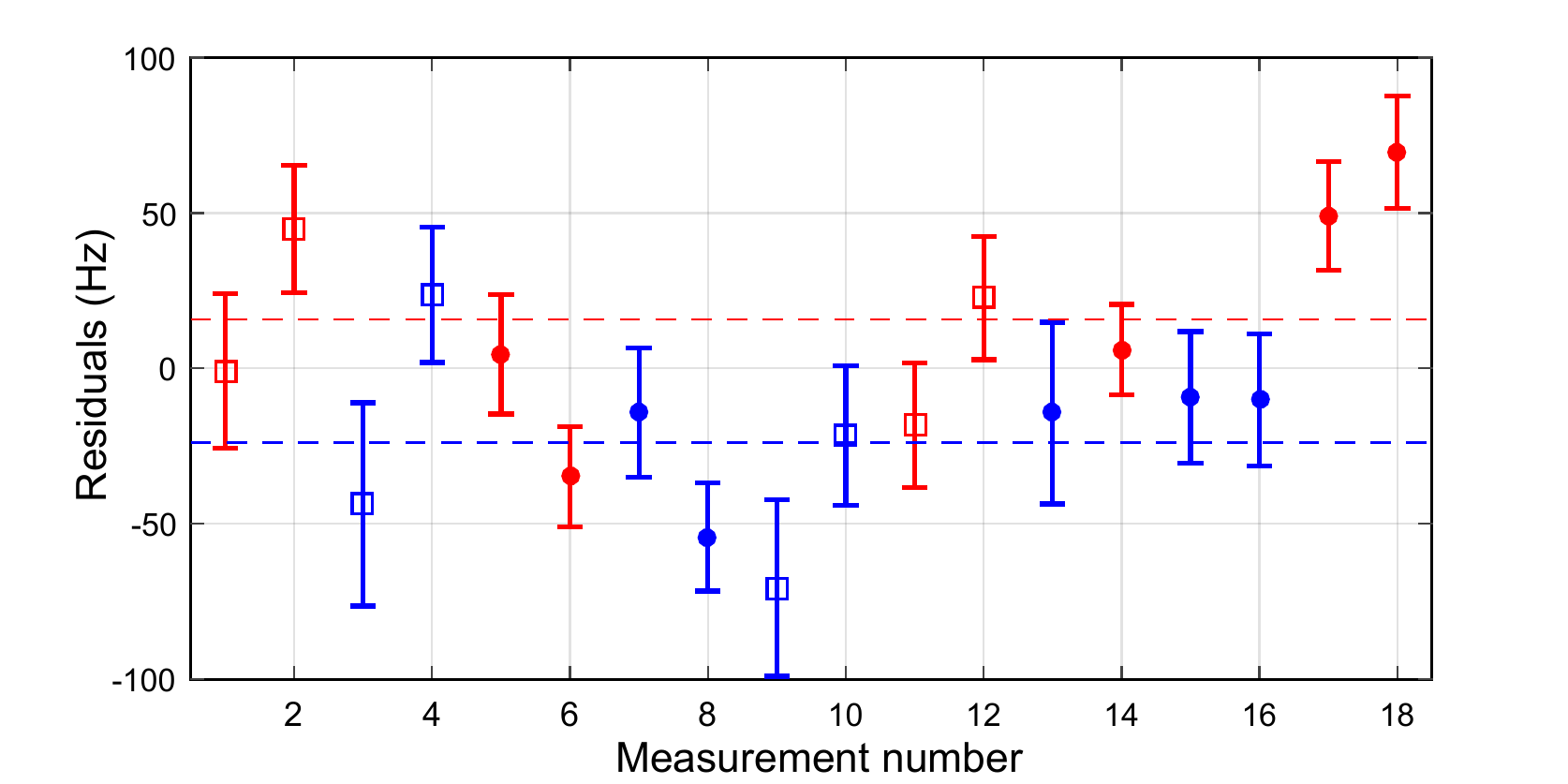}
\caption{\label{fig:residuals}
Residuals of the 18 frequency measurements from the linear fit. Data points 1-8 were taken with the frequency comb's repetition rate locked to a stable quartz oscillator. Data points 9-18 were recorded with a comb line locked to the frequency of the laser at 729~nm. Measurements of the $m=5/2\leftrightarrow m=7/2$ transition are denoted by squares, measurements of the $m=-5/2\leftrightarrow m=-7/2$ transition are denoted by circles. The colour indicates whether the duration of the Ramsey experiments was 100$\mu$s (blue) or 200$\mu$s (red).
Dashed lines indicate the means of the residuals with the corresponding colour.
}
\end{center}
\end{figure}

We plot the measured frequency $f_i$ as a function of $s_\pm B$ and fit a straight line to the data in order to extract $f_0$ and the g-factor of the $^3\mathrm{P}_1,F^\prime=7/2$ state. This procedure yields a transition frequency at $B=0$ of $f_0=1122\,842\,857\,334\,711$~Hz with a statistical uncertainty of $36$~Hz. The residuals of the fit are shown in Fig.~\ref{fig:residuals} with error bars accounting for the standard deviation of the fifty frequency measurements per data set and errors in the determination of the magnetic field. 

The scatter of the data points is somewhat bigger than what one would expect given the error bars of the individual data points. Moreover, the measurements with free Ramsey durations of 200~$\mu$s yield a frequency estimate differing by $\delta\nu=40$~Hz from the measurement based on Ramsey experiments of 100~$\mu$s duration. To evaluate the statistical significance of this finding, we test the null hypothesis $H$ that the determined frequency is independent of the duration of the Ramsey experiment. Under this hypothesis, the difference $\Delta = 2/N\sum_{i=1}^{N/2}(X_i^{(200)}-X_i^{(100)})$ (with ${\cal N}(0,\sigma_r^2)$-distributed Gaussian random variable $X_i^{(T)}$ describing the outcome of the $i^{th}$ measurement with Ramsey duration $T$) is itself a Gaussian random variable with a distribution given by ${\cal N}(0,\sigma^2)$ with $\sigma=\sqrt{4/N}\sigma_r=17$~Hz. Then, the probability of finding a difference of $\delta$ or larger is given by $p=P(|\Delta|\ge|\delta||H)=1-\mbox{erf}(\delta/(\sqrt{2}\sigma))\approx 0.02$. This rather small $p$-value makes it seem unlikely that there was no systematic shift present in the experiment. However, no such shift was detected in subsequent control experiments probing the transition by alternating between Ramsey experiments of 100 and 200 $\mu$s duration.
From the variation of the magnetic-field-induced splitting $\Delta \nu/B = \frac{\mu_B}{h}({\textstyle\frac{7}{2}}g_{^{\,3\!}P_1,F^\prime=7/2}-{\textstyle\frac{5}{2}}g_{^{\,1\!}S_0}) = 2.100056(9)$\,MHz$/$G and the knowledge of $g_{^{\,1\!}S_0}$~\cite{Rosenband:2007}, we determine the Land\'e g-factor of the excited state to be $g_{^{\,3\!}P_1,F^\prime=7/2} = 0.428132(2)$. 
 
\subsection{Evaluation of systematic frequency shifts and measurement uncertainties} \label{section:systematic_shifts}

For the evaluation of the $^1\mathrm{S}_0\leftrightarrow{^3\mathrm{P}_1}$ transition frequency, both systematic frequency shifts in Al$^+$ and Ca$^+$ need to be taken into consideration. The scatter of the frequency measurements shown in Fig.~\ref{fig:residuals} suggests that an unexpected systematic shift might have been present when these measurements were taken. Even though it seems unlikely, the frequency shift of 40~Hz between the measurements taken with 100~$\mu$s (200~$\mu$s) Ramsey time could have been caused by an ac-Stark shift shifting the transition frequency during the $\pi/2$-pulses. This effect would vanish in the limit of infinitely long Ramsey experiments and lead to a systematic shift of 85~Hz in the frequency measurement. As we cannot exclude the existence of such a shift, we added an uncertainty of 85~Hz to our error budget.  

The electric quadrupole moment of the $^{3}\mathrm{P}_1$ level of the $^{27}\mathrm{Al}{^+}$ ion couples to any static electric field gradients in the trap. The dominant electric field gradient is caused by the quadrupolar static electric potential confining ions axially and by the electric field gradient of the co-trapped $^{40}\mathrm{Ca}^+$ ion. These field gradients will shift the energy of the stretched $^{3}\mathrm{P}_1,F=7/2,m_F=\pm7/2$ levels by $\Delta E=\frac{1}{2\sqrt{30}}\langle nJ||Q_2||nJ\rangle\frac{\partial^2\Phi}{\partial z^2}$. Following the prescription of Refs. \cite{Beloy:2017,Wuebbena:2012} for calculating the second derivative of the potential for a mixed ion crystal with 888~kHz axial in-phase oscillation frequency, one obtains a frequency shift of -7.4~Hz for the stretched Zeeman states where we used the reduced matrix element $\langle nJ||Q_2||nJ\rangle$ calculated in Ref.~\cite{Beloy:2017} for the estimation of the shift.

The fine-structure mixing effect of a non-zero magnetic field gives rise to a small second-order Zeeman shift of the $|^{3}\mathrm{P}_1,F=7/2,m_F=\pm7/2\rangle$ states via a coupling to the $^{3}\mathrm{P}_2$ states with the same magnetic quantum number. Using second order perturbation theory, the quadratic Zeeman shift can be approximated by
$\Delta E^{(2) Z}_{J,M}(^{3}P_{1})=-\frac{1}{2}\zeta_{nLS}(B/B_{\rm{fs}})^2$,
\cite{JookWalraven2018, sobelman2012atomic} where $B_{\rm{fs}}\equiv \zeta_{nLS}J'/\mu'_{B}$ is the fine-structure crossover field, with $\mu'_{B}\equiv (g_{s}-g_{J})\mu_{B}\cong \mu_{B}$, and $\zeta_{nLS}=1.8591$\,THz the coupling constant. For the $^3$P$_{2}$-$^3$P$_{1}$ fine-structure splitting we have L=1, S=1, J'= 2 and B$_{\rm{fs}}$=265.344$ \times 10^{4}$\,G. We calculate for the quadratic frequency shift of the J=M=1 level that $d^2\nu/dB^2=-263.74 \,\mathrm{mHz/G}^2$. For a magnetic field of approximately 4\,G the systematic shift amounts to only -2.109(0.005)\,Hz.

In $^{40}\mathrm{Ca}^+$, the $D_{5/2}$ state has an electric quadrupole moment of $\Theta(D,5/2)=1.83(1)\,ea_0^2$ \cite{Roos:2006} where $e$ is the elementary charge and $a_0$ the Bohr radius. The interaction of the electric quadrupole moment with the static electric field gradient of the linear ion trap (i.e. the dominant field gradient in our experiments) upshifts the $m=\pm 3/2$ Zeeman states by 2.83~Hz. Additionally, there is a second-order Zeeman shift by the magnetic field that couples the Zeeman levels of the $\mathrm{D}_{5/2}$ state to the levels of the $\mathrm{D}_{3/2}$ state having the same magnetic quantum number. A field of 4~G causes a frequency shift of the $\mathrm{D}_{5/2},m=\pm3/2$ levels of 2.79~Hz. Both shifts taken together result in a 15.1(3)~Hz correction of the transition frequency in Al$^+$. Other sources of frequency shifts such as black-body radiation, time-dilation shifts by excess micromotion or non-zero secular motion are expected to be no bigger than 1~Hz. This claim is supported by the absence of any large frequency shifts in the comparison with other Ca$^+$ experiments (subsection~\ref{section:ProbingCa}). 

The systematic frequency shifts discussed above are summarised in Table \ref{table:shifts}. Taking them into consideration leads to a correction of 24.5~Hz of the transition frequency and results in a measured transition frequency of $\nu_\mathrm{Al}= 1122\,842\,857\,334\,736(93)$~Hz.

\begin{table}
\begin{tabular}{lcc}
 Effect & Ca shift (uncertainty)[Hz]  & Al shift (uncertainty) [Hz]  \\
 \hline\hline
 $S_{1/2}-D_{5/2}$ frequency & 0 (4)  & - \\
 Electric quadrupole shift& 2.7(0.1)  & -7.4 \\
 2nd order Zeeman shift&  2.8 & -2.1  \\
 Ramsey probe time dependence& - & 0 (85) \\
 Statistics& - & 0 (36) \\
 \hline
 Total shift& 5.5(4.0)  & -9.5 (92.0)
\end{tabular}
\caption{Systematic frequency shifts and measurement uncertainties of the two transitions probed in $\mathrm{Ca}^+$ and $\mathrm{Al}^+$. For the determination of the intercombination line frequency, the systematic shifts of the calcium transition enter the error budget after having been multiplied by the frequency ration $\nu_{\mathrm{Al}}/\nu_{\mathrm{Ca}}$.}
\label{table:shifts}
\end{table}

Systematic frequency shifts also need to be considered for the determination of the $^3\mathrm{P}_1$ state's g-factor. Electric quadrupole shifts and second-order Zeeman shifts do not affect the g-factor determination as they shift both stretched Zeeman states by the same amount. Ac-Zeeman and ac-Stark shifts, however, need to be considered. The ac-magnetic field generated by the ion trap drive at 32~MHz gives rise to a shift of the probed $|\mathrm{S}_{1/2},m=\pm 1/2\rangle\leftrightarrow|\mathrm{D}_{5/2},m=\pm3/2\rangle$ Zeeman transitions in Ca$^+$ that mimics an additional dc-magnetic field of about 4~$\mu$G. For the dc-field of 4~G used in the experiment, this corresponds to a fractional error of $10^{-6}$ in the g-factor determination which is below our measurement uncertainty. In Al$^+$, due to the smaller splitting of the Zeeman levels, the ac-magnetic field shift is even smaller. The static magnetic field gradient of 30~$\mu$G/m causes only a negligible magnetic field difference in the two ion locations and can thus be neglected. Ac-Stark shifts on the $^1\mathrm{S}_0\leftrightarrow{^3\mathrm{P}_1}$ transition that arise from off-resonant coupling to the axial out-of-phase sideband transitions or to $\Delta m=0,\pm 1$ transitions induce fractional errors in the g-factor determination of well below $10^{-6}$. Therefore, we conclude that systematic shifts do not alter the measured g-factor value of $g_{^{\,3\!}P_1,F^\prime=7/2} = 0.428132(2)$.

\begin{figure}[tb]
\begin{center}
\includegraphics[height=0.5\textwidth]{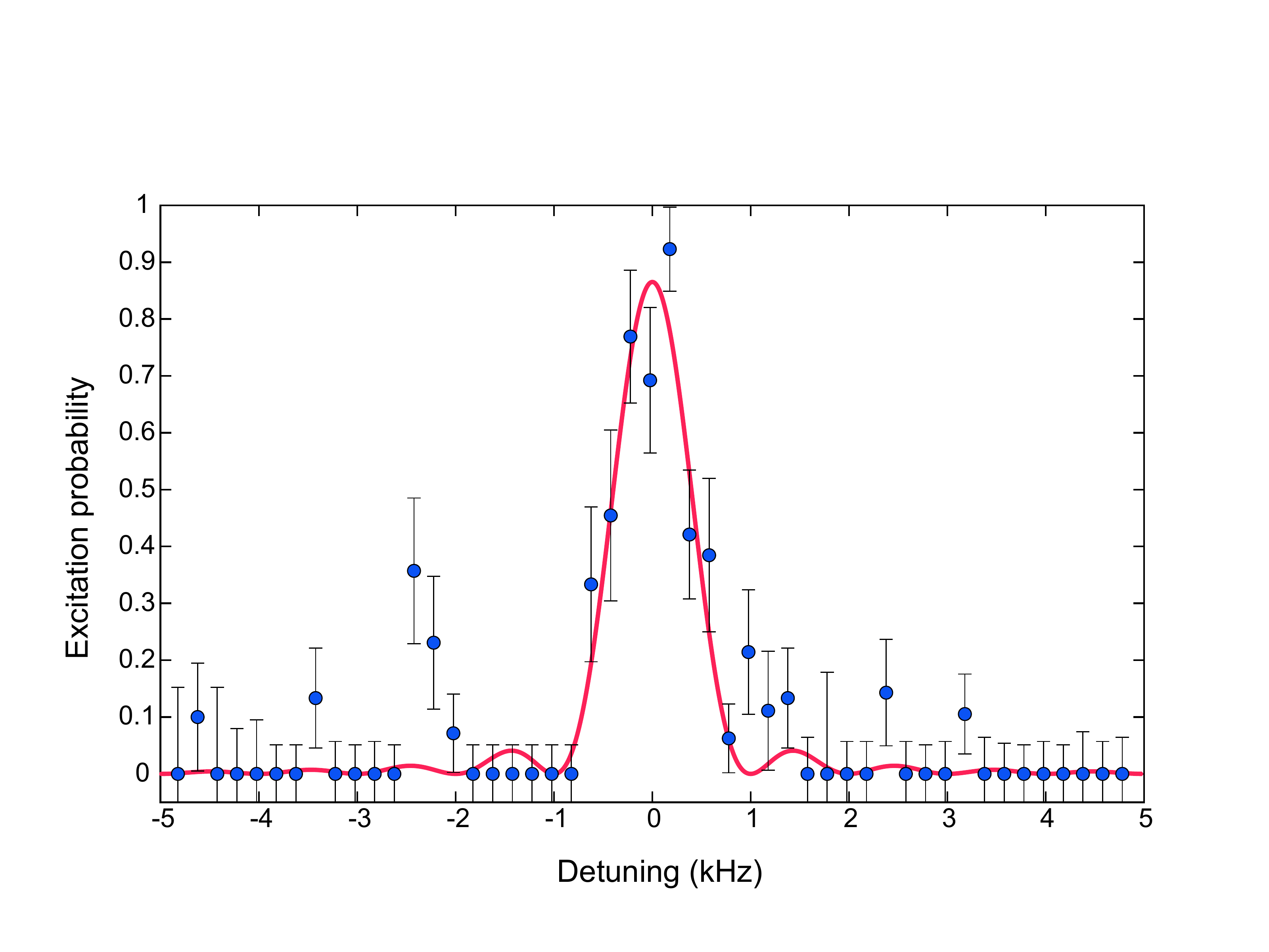}
\caption{\label{fig:clocktransition}
Excitation of the $|^1\mathrm{S}_0,m_F=-5/2\rangle\leftrightarrow|^3\mathrm{P}_0,m_{F^\prime}=-3/2\rangle$ transition with a laser pulse of 1~ms duration. The excitation probability is inferred by measuring by the probability of the ion to undergo a change of its electronic state from one experiment to the next.
For probing the transition, the measurement protocol described in subsection \ref{section:QLS} is modified by probing the clock transition instead of the intercombination line in step (iv) and repeating the quantum logic steps (v) and (vi) twice. In this way, errors arising from imperfect state mapping from Al$^+$ to Ca$^+$ are suppressed.
}
\end{center}
\end{figure}

\section{Discussion and conclusion} \label{conclusion}
Using the S$_{1/2}\leftrightarrow$D$_{5/2}$ quadrupole transition of $^{40}$Ca$^{+}$ as a frequency reference, we have employed quantum logic spectroscopy for a measurement of the (3s$^{2}$)$^{1}$S$_{0}\leftrightarrow (3$s3p)$^{3}$P$_{1}$, F =7$/$2 intercombination transition in $^{27}$Al$^{+}$ with an uncertainty of \~100\,Hz. Our results are consistent with previous measurements of the transition \cite{Rosenband:2005} and improve the uncertainty by one order of magnitude. While the transition itself is not particularly interesting as a frequency standard, its coherent excitation is an enabling technology for detecting population in the $^3\mathrm{P}_0$ state when exciting the highly relevant $^1\mathrm{S}_0\leftrightarrow{^3\mathrm{P}_0}$ clock transition.  

Recently, we have also probed the $^1\mathrm{S}_0\leftrightarrow^3\mathrm{P}_0$ clock transition. Figure~\ref{fig:clocktransition} shows a spectrum of the $(m=-5/2)\leftrightarrow(m=-3/2)$ Zeeman transition with a probe-time limited width of 1~kHz. Efforts to obtain spectrally narrower lines are hampered by two experimental shortcomings, namely frequency drifts of the frequency-quadrupled fibre laser probing the clock transition and creation of molecular ions by reaction of $\mathrm{Al}^+$ with $\mathrm{H}_2$ background gas molecules. Molecular ion creation limits the time for experiments probing the clock transition to about 15~minutes, which, in combination with long $\mathrm{Al}^+$ reloading times \cite{Guggemos:2015}, prevented us from investigating the clock transition more thoroughly. This will be subject to future work once the apparatus has been improved.

\ack We acknowledge funding by the European Space Agency (contract number 4000102396/10/NL/SFe) and by the European Commission via the integrated project SIQS, and by the Institut f\"ur Quanteninformation GmbH. We would also like to thank J. Walraven and K. Beloy for helpful insights in the calculation of systematic shifts.

\section*{References}

\bibliography{ref}

\providecommand{\newblock}{}
\begin{thebibliography}{10}
\expandafter\ifx\csname url\endcsname\relax
  \def\url#1{{\tt #1}}\fi
\expandafter\ifx\csname urlprefix\endcsname\relax\def\urlprefix{URL }\fi
\providecommand{\eprint}[2][]{\url{#2}}

\bibitem{Dehmelt:1982}
Dehmelt H~G 1982 {\em IEEE Trans.~Instr.~Meas.\/} {\bf 31} 83--87

\bibitem{Ludlow:2015}
Ludlow A~D, Boyd M~M, Ye J, Peik E and Schmidt P~O 2015 {\em Rev. Mod. Phys.\/}
  {\bf 87} 637--701

\bibitem{Nemitz:2016}
Nemitz N, Ohkubo T, Takamoto M, Ushijima I, Das M, Ohmae N and Katori H 2016
  {\em Nat. Phot.\/} {\bf 10} 258

\bibitem{Schioppo:2017}
Schioppo M, Brown R~C, McGrew W~F, Hinkley N, Fasano R~J, Beloy K, Yoon T~H,
  Milani G, Nicolodi D, Sherman J~A, Phillips N~B, Oates C~W and Ludlow A~D
  2017 {\em Nat. Phot.\/} {\bf 11} 48--52

\bibitem{Oelker:2019A}
Oelker E, Hutson R~B, Kennedy C~J, Sonderhouse L, Bothwell T, Goban A, Kedar D,
  Sanner C, Robinson J~M, Marti G~E, Matei D~G, Legero T, Giunta M, Holzwarth
  R, Riehle F, Sterr U and Ye J 2019 {\em arXiv:1902.02741\/}

\bibitem{Brewer:2019A}
Brewer S~M, Chen J~S, Hankin A~M, Clements E~R, Chou C~W, Wineland D~J, Hume
  D~B and Leibrandt D~R 2019 {\em arXiv:1902.07694\/}

\bibitem{Huntemann:2016}
Huntemann N, Sanner C, Lipphardt B, Tamm C and Peik E 2016 {\em Phys. Rev.
  Lett.\/} {\bf 116}(6) 063001

\bibitem{Chen:2017}
Chen J~S, Brewer S~M, Chou C~W, Wineland D~J, Leibrandt D~R and Hume D~B 2017
  {\em Phys. Rev. Lett.\/} {\bf 118}(5) 053002

\bibitem{hankin2019systematic}
Hankin A, Clements E, Huang Y, Brewer S, Chen J~S, Chou C, Hume D and Leibrandt
  D 2019 {\em arXiv preprint arXiv:1902.08701\/}

\bibitem{Rosenband:2008}
Rosenband T, Hume D~B, Schmidt P~O, Chou C~W, Brusch A, Lorini L, Oskay W~H,
  Drullinger R~E, Fortier T~M, Stalnaker J~E, Diddams S~A, Swann W~C, Newbury
  N~R, Itano W~M, Wineland D~J and Bergquist J~C 2008 {\em Science\/} {\bf 319}
  1808--1812

\bibitem{Oskay:2006}
Oskay W~H, Diddams S~A, Donley E~A, Fortier T~M, Heavner T~P, Hollberg L, Itano
  W~M, Jefferts S~R, Delaney M~J, Kim K, Levi F, Parker T~E and Bergquist J~C
  2006 {\em Phys. Rev. Lett.\/} {\bf 97} 020801

\bibitem{Barwood:2014}
Barwood G~P, Huang G, Klein H~A, Johnson L~A~M, King S~A, Margolis H~S,
  Szymaniec K and Gill P 2014 {\em Phys. Rev. A\/} {\bf 89}(5) 050501

\bibitem{Huang:2016}
Huang Y, Guan H, Liu P, Bian W, Ma L, Liang K, Li T and Gao K 2016 {\em Phys.
  Rev. Lett.\/} {\bf 116}(1) 013001

\bibitem{Rosenband:2006}
Rosenband T, Itano W~M, Schmidt P~O, Hume D~B, Koelemeij J~C~J, Bergquist J~C
  and Wineland D~J 2006 Blackbody radiation shift of the $^{27}${A}l$^+$
  $^1${S}$_0\rightarrow^3${P}$_0$ transition {\em Proceedings of the 20th
  European Frequency and Time Forum\/}

\bibitem{Beloy:2017}
Beloy K, Leibrandt D~R and Itano W~M 2017 {\em Phys. Rev. A\/} {\bf 95}(4)
  043405

\bibitem{Schmidt:2005}
Schmidt P~O, Rosenband T, Langer C, Itano W~M, Bergquist J~C and Wineland D~J
  2005 {\em Science\/} {\bf 309} 749--752

\bibitem{Wuebbena:2014}
Wuebbena J 2014 {\em Controlling Motion in Quantum Logic Clocks ({PhD} thesis,
  {L}eibniz {U}niversit\"at {H}annover)\/} phdthesis Leibniz Universit\"at
  Hannover

\bibitem{Xu:2016}
Xu Z~T, Yuan W~H, Zeng X~Y, Che H, Shi X~H, Deng K, Zhang J and Lu Z~H 2016
  {\em Journal of Physics: Conference Series\/} {\bf 723} 012026

\bibitem{Cui:2018}
Cui K~F, Shang J~J, Chao S~J, Wang S~M, Yuan J~B, Zhang P, Cao J, Shu H~L and
  Huang X~R 2018 {\em Journal of Physics B: Atomic, Molecular and Optical
  Physics\/} {\bf 51} 045502

\bibitem{Hannig:2019}
Hannig S, Pelzer L, Scharnhorst N, Kramer J, Stepanova M, Xu Z~T, Spethmann N,
  Leroux I~D, Mehlst\"aubler T~E and Schmidt P~O 2019 {\em arXiv:1901.02250\/}

\bibitem{kramida2018nist}
Kramida A, Ralchenko Y, Reader J, NIST and {ASD Team} 2018 {\textit{NIST Atomic
  Spectra Database (version 5.6.1)}} National Institute of Standards and
  Technology, Gaithersburg, MD. DOI: https://doi.org/10.18434/T4W30F

\bibitem{Wineland:2002}
Wineland D, Bergquist J, Bollinger J, Drullinger R and Itano W 2002 Quantum
  computers and atomic clocks {\em Proc. 6th Symposium on Frequency Standards
  and Metrology, St. Andrews, Scotland, Sept. 9 - 14, 2001.\/} ed Gill P
  (Singapore: World Scientific) pp 361--368

\bibitem{Rosenband:2007}
Rosenband T, Schmidt P, Hume D, Itano W, Fortier T, Stalnaker J, Kim K, Diddams
  S, Koelemeij J, Bergquist J and Wineland D 2007 {\em Phys. Rev. Lett.\/} {\bf
  98} 220801

\bibitem{Hempel:2013}
Hempel C, Lanyon B~P, Jurcevic P, Gerritsma R, Blatt R and Roos C~F 2013 {\em
  Nat. Phot.\/} {\bf 7} 630--633

\bibitem{Wan:2014}
Wan Y, W\"ubbena F~G~B, Scharnhorst N, Leroux S~A~D, L\"orch B~H, Hammerer K
  and Schmidt P~O 2014 {\em Nat. Comm.\/} {\bf 5} 4096

\bibitem{Wolf:2016}
Wolf F, Wan Y, Heip J~C, Gebert F, Shi C and Schmidt P~O 2016 {\em Nature\/}
  {\bf 530} 457--460

\bibitem{Chou:2017}
Chou C~W, Kurz C, Hume D~B, Plessow P~N, Leibrandt D~R and Leibfried D 2017
  {\em Nature\/} {\bf 545} 203--207

\bibitem{Traebert:1999}
Tr\"abert E, Wolf A, Linkemann J and Tordoir X 1999 {\em Journal of Physics B:
  Atomic, Molecular and Optical Physics\/} {\bf 32} 537

\bibitem{Hume:2007}
Hume D~B, Rosenband T and Wineland D~J 2007 {\em Phys.~Rev.~Lett.\/} {\bf 99}
  120502

\bibitem{Rosenband:2005}
Rosenband T, Schmidt P~O, Kobayashi Y, Langer C, Itano W~M, Diddams S~A,
  Bergquist J and Wineland D~J 2005 Al$^+$ spectroscopy via sympathetic cooling
  and quantum information transfer using {B}e$^+$. {\em Abstract submitted for
  the {DAMOP}05 meeting of the {A}merican {P}hysical {S}ociety.
  \url{http://meetings.aps.org/link/BAPS.2005.DAMOP.L4.7}\/}
  \urlprefix\url{http://meetings.aps.org/link/BAPS.2005.DAMOP.L4.7}

\bibitem{Wuebbena:2012}
W\"ubbena J~B, Amairi S, Mandel O and Schmidt P~O 2012 {\em Phys. Rev. A\/}
  {\bf 85} 043412

\bibitem{Guggemos:2015}
Guggemos M, Heinrich D, Herrera-Sancho O~A, Blatt R and Roos C~F 2015 {\em New
  J. Phys.\/} {\bf 17} 103001

\bibitem{Ma:1994}
Ma L, Jungner P, Ye J and Hall J~L 1994 {\em Opt.~Lett.\/} {\bf 19} 1777

\bibitem{Chwalla:2009}
Chwalla M, Benhelm J, Kim K, Kirchmair G, Monz T, Riebe M, Schindler P, Villar
  A~S, H\"ansel W and Roos C~F 2009 {\em Phys.~Rev.~Lett.\/} {\bf 102} 023002

\bibitem{Colombe:2014}
Colombe Y, Slichter D~H, Wilson A~C, Leibfried D and Wineland D~J 2014 {\em
  Optics Express\/} {\bf 22} 19783

\bibitem{Roos:2008a}
Roos C~F, Monz T, Kim K, Riebe M, H\"affner H, James D~F~V and Blatt R 2008
  {\em Phys.~Rev.~A\/} {\bf 77} 040302(R)

\bibitem{Matsubara:2012}
Matsubara K, Hachisu H, Li Y, Nagano S, Locke C, Nogami A, Kajita M, Hayasaka
  K, Ido T and Hosokawa M 2012 {\em Opt. Express\/} {\bf 20} 22034--22041

\bibitem{JookWalraven2018}
Walraven J 2018 {Private Communication}

\bibitem{sobelman2012atomic}
Sobelman I~I 2012 {\em Atomic spectra and radiative transitions\/} vol~12
  (Springer Science \& Business Media)

\bibitem{Roos:2006}
Roos C~F, Chwalla M, Kim K, Riebe M and Blatt R 2006 {\em Nature\/} {\bf 443}
  316--319

\end{thebibliography}

\end{document}